# Naturally Resonant Emitters: Approaching Fundamental Antenna Limits

Damir Latypov

CRG Defense, 8821 Washington Church Rd., Miamisburg. OH 45342

latypovdm@CrgDefense.com

*Abstract*

Antenna miniaturization remains a critical technological challenge across frequency scales – from microwave RF links in phones/wearables to VLF/ELF for underwater-to-air communications and ionospheric probing. At deeply subwavelength scales ($ka \ll 1$), conventional antennas require complex and lossy matching circuits due to absent intrinsic material resonances, motivating resonant electrically small emitters (ESEs) like mechanical resonators and quantum emitters. Here, we extend the theory of electrically small antennas (ESAs) to this broader ESE class, deriving the fundamental efficiency limit for a unit volume emitter at given frequency and bandwidth. Our figure of merit (FOM) – quantifying proximity to this limit – enables direct comparison across ESE types, frequencies, bandwidths and scales. We demonstrate its utility using public data from ELF (decommissioned) and VLF Navy facilities alongside two mechanical ESEs reported in literature. The measurements reveal that mechanical antennas operate near theoretical FOM limit, questioning claims of possible further orders-of-magnitude gains. A naturally resonant emitter is still subject to the Chu-Harrington limit (CHL) under its standard assumptions. Indeed, we derive novel CHL-dictated constraints on atomic ESE properties: lower bound on excited-state lifetime and an upper bound on transition dipole moment. Classical CHL circumvention strategies – like non-Foster matching and nonlinear parametric antennas – have been researched for decades. Quantum degrees of freedom, such as superradiance via collective coherence represent the new frontier.

I. Introduction

Antenna miniaturization is a critical technological challenge across frequency scales. At microwave frequencies, it enables reliable RF links to fit into phones, wearables, implants, and dense IoT nodes where volume, weight, and cost are tightly constrained. At much lower VLF and ELF frequencies, compact antennas would enable underwater-to-air RF links, ionospheric/magnetospheric probing from space platforms, geological explorations, etc.

Miniaturization amounts to operation at deeply subwavelength scales ($ka \ll 1$), where $a$ is a characteristic antenna length and $k$ is the free-space wavenumber. Conventional electromagnetic antennas, which are distributed structures without intrinsic material resonances, cannot resonate naturally in this regime. This limitation has motivated exploration of alternative RF emitters, such as mechanical resonators [1], [2], and quantum emitters [3], [4], [5], that exhibit natural resonances at those scales.

For these novel devices, the terminology warrants reexamination: should they be called antennas? The IEEE Std 145 defines antenna as "that part of transmitting or receiving system that is designed to radiate or receive electromagnetic waves" [6]. The same document also states, "It is assumed in this standard that an antenna is a passive linear reciprocal device". In the small-signal regime without active modulation, mechanical resonators satisfy this definition. The quantum emitters, however, do not since they are not generally reciprocal [3], [4].

In this paper, we focus on radiators (not receivers) and adopt the term "emitter" to encompass the quantum devices as well; we use "antenna" interchangeably when IEEE definition holds. Specifically, for our electrically small ($ka < 1$) focus, we employ ESE (electrically small emitter) as generalization of the standard term ESA (electrically small antenna).

Now that ESA has been expanded to ESE, how do we compare ESEs? Conventional electrically small antennas (ESAs) encounter fundamental bandwidth-efficiency trade-offs due to reactive near-field dominance, as quantified by the Chu-Harrington limit (CHL) [7]. Remarkably, the CHL itself is agnostic to the specific resonance mechanisms – it arises from the minimum stored electromagnetic energy that any classical time-harmonic source must confine within the smallest enclosing sphere independent of tuning, matching or excitation method [8]. Thus, CHL-compliant emitters differ primarily in the costs of achieving the resonance – such as associated size constraints and non-radiative losses. To capture these, and enable ESE comparison across frequencies and bandwidth, we establish an ESE figure of merit (FOM) which quantifies its deviation from the CHL-derived efficiency bound.

Besides the ESE's FOM comparison metric, this paper presents two more contributions. First, we derive a fundamental upper bound for the radiation power density (at a fixed bandwidth) for any CHL-compliant ESE. Second, we leverage CHL universality to establish limits on ESE properties - specifically, bounds on the excited state lifetime and transition dipole moment of an atom.

The paper is organized as follows. Section II provides background on CHL. Section III introduces the ESEs FOM comparison metric and derives a fundamental upper bound on radiated power density (at fixed bandwidth, per unit of input power). Section IV examines quantum effects within the CHL, establishing a lower bound on the excited state lifetime and an upper bound the transition dipole moment of an atom. Section V provides conclusions.

## II.     Preliminaries

The stored-energy quality factor $Q$ of an antenna is defined as

$$Q_{stored} = \frac{\omega(W_m + W_e)}{P_{diss}}, \qquad (1)$$

where $\omega$ is the angular frequency, $W_m$ and $W_e$ are the time-averaged stored magnetic and electric energies, and $P_{diss}$ is the dissipated power which includes the radiated power and other losses such as Ohmic and dielectric:

$$P_{diss} = P_{rad} + P_{loss}. \qquad (2)$$

For a lossless, passive, time-invariant, linear antenna operating in free space, $Q_{stored}$ is bounded from below by the Chu-Harrington limit (CHL) [7]:

$$Q_{stored} \geq Q_{CHL} = \frac{1}{(ka)^3} + \frac{1}{ka}, \qquad (3)$$

where $a$ is the radius of the smallest sphere enclosing the antenna, and $k = 2\pi/\lambda$ is the free-space wavenumber. Although derived specifically for ESAs, CHL is applicable to any classical ESEs whenever these assumptions hold. This is evident from the field-based derivation of the CHL presented by Collin and Rothschild in [8].

For quantum emitters, one should consider three scenarios: a single quantum emitter, an incoherent macroscopic ensemble of quantum emitters, and a coherent ensemble. The expression for radiated power due to an atomic transition is identical to that for a classical dipole up to a numerical factor - a quantum emitter could be twice as bright as its classical counterpart due to the quantum vacuum fluctuations [9]. Thus, it is reasonable to hypothesize that quantum emitter is subject to CHL. In case of incoherent macroscopic ensembles of quantum emitters, the Bohr correspondence principle dictates that the radiation should become effectively classical, so CHL apply. In a coherent ensemble, the collective, entangled nature of the emission could produce superlinear power scaling with number of emitters (superradiance). A higher radiated power at fixed stored energy corresponds to a shorter radiation pulse and therefore a larger radiation bandwidth. This suggests that a coherent macroscopic ensemble of quantum emitters can overcome the technological constraints imposed by CHL, enabling higher bandwidth and efficiency than would be possible in a classical antenna of the same electrical size.

Near a single, simple resonance (Lorentzian-like response, second-order behavior), $Q_{stored} = Q_{bw}$, where the bandwidth quality factor is

$$Q_{bw} = \frac{f}{\Delta f}. \tag{4}$$

Here $f$ is the resonant frequency, and $\Delta f$ is the 3-dB bandwidth.

Thus, the CHL bounds $Q_{stored}$ directly via (1), and bounds achievable radiation bandwidth when $Q_{stored} = Q_{bw}$, (under narrowband, single-resonance conditions). Hereafter, unless stated otherwise, we assume this equivalence holds and denote both $Q$-factors simply as $Q$.

### III. CHL and radiation efficiency

What does CHL tell us about the radiation efficiency of ESA? Radiation efficiency $\eta$ for an impedance-matched antenna is given by

$$\eta = \frac{P_{rad}}{P_{rad} + P_{loss}}. \tag{5}$$

CHL is derived for lossless antennas, so a strictly CHL-compliant antenna has $\eta$=1. Several papers investigated the upper bounds on radiation efficiency due to finite conductivity of metallic antennas, see e.g., [10], [11]. In deep subwavelength regime ($ka \ll 1$, high $Q$) which is of primary interest in this paper, the radiation efficiency is bounded not by the finite conductivity but by a minimum bandwidth requirement. In passive systems, the only way to exceed the CHL on bandwidth (under single-mode $Q_{stored} = Q_{bw}$ assumption) is to add loss, as seen in (1), (2). Loss reduces the lower bound on bandwidth quality factor (4) from $Q_{bw} \geq Q_{CHL}$ to $Q_{bw} \geq \eta Q_{CHL}$. This leads to a bound on radiating efficiency for a given bandwidth:

$$\eta \leq \min\left(\frac{f}{\Delta f}\frac{1}{Q_{CHL}}, 1\right). \tag{6}$$

To our knowledge, this CHL-derived efficiency bound has not been previously reported. We illustrate it using two examples: the ELF facility at Clam Lake, Wisconsin (decommissioned in 2004), and the VLF facility at Cutler, Maine.

The ELF facility at Clam Lake used two orthogonal 14-mile-long dipoles, so the radius of the enclosing sphere is $14/\sqrt{2}$ miles. It employed minimum-shift keying frequency modulation, where the signal consisted of smoothly connected segments of sinusoids with frequencies of 72 Hz and 80 Hz, with the center frequency of 76 Hz. The measured power spectrum of this signal reported in [12], matches a single simple resonance model validating the $Q_{stored} = Q_{bw}$ equivalence. The facility generated up to 2.3 MW of power [13]; at typical 1 MW input, it radiated ~2 W of ELF signal which gives radiating efficiency $\eta \sim 2 \cdot 10^{-6}$. The upper CHL-bound on efficiency given by (6) is $1.5 \cdot 10^{-4}$.

The VLF facility at Cutler has rated power of more than 2 MW and currently operates at 24 kHz (call signal NAA). Antenna array geometry data needed to compute the enclosing sphere can be found in [14]. The bandwidth of 240 Hz is estimated from the Navy-specified 200 VLF baud rate, using the 3dB bandwidth formula $BW \approx 1.2 \times baud\ rate$ [15]. Unlike Clam Lake ELF, the Cutler VLF transmitter NAA's power spectrum does not appear in publicly available data. However, since $f/\Delta f = 100$ exceeds $Q_{CHL} \approx 11.7$, no additional losses are needed to achieve the 240 Hz bandwidth. As the result, the NAA transmitter has high efficiency.

Table 1 summarizes these CHL efficiency bound calculations for both facilities.

Table 1. Efficiency bound calculation for ELF facility at Clam Lake and VLF facility at Cutler

| Facility | $f$(Hz) | $\Delta f$(Hz) | $a$(m) | $Q_{CHL}$ | Efficiency bound (6) | Reported efficiency |
|---|---|---|---|---|---|---|
| ELF | 76 | 4 | 15932 | $6.1 \cdot 10^4$ | $1.5 \cdot 10^{-4}$ | $\sim 2 \cdot 10^{-6}$ |
| VLF | 24000 | 240 | 935 | 11.7 | 1 | >0.5 |

Now we apply eq. (6) to two mechanical antennas from literature. Ref [1] reports a 9.4 cm long, 1.6 cm diameter lithium niobite (LN) piezoelectric rod with resonant frequency 35.568 kHz and measured total $Q$=303,000. Assuming VSWR=2, the authors estimated natural bandwidth $\Delta f$=84 mHz. With these parameters, the efficiency bound (6) is $1.82 \times 10^{-8}$. This falls within the author's efficiency estimate in the range $2 \times 10^{-7}$ and $1 \times 10^{-8}$. The higher end reflects a secondary resonance downshifted by 7 Hz from the primary, where the total $Q$ reaches 615,000 - doubling the efficiency achieved at the primary frequency.

In [2], the mechanical antenna comprises a lead zirconate titanate (PZT) disk (8 cm diameter, 1 cm height) modulated via binary frequency phase-shift keying (BFSK) within its 3-dB bandwidth at $f_1$=33.218 kHz and $f_2$=33.248 kHz. Input power of 1.2 W excites the resonator. The authors claim that due to the high relative permittivity of PZT ($\varepsilon_r > 1000$), their antenna reaches its far-field regime after only 1.3 m. The measured magnetic field $B$=50 fT (RMS, ±10 fT uncertainty) at distance $R$=4.5 m from the antenna (Figure 4C of [2]). For an isotropic radiator this corresponds to radiated power

$$P_{rad} = \frac{c}{2\mu_0} B^2 \times 4\pi R^2 \sim 78.9 \pm 30.4\ pW. \qquad (7)$$

Since the mechanical antenna behaves as an electrically small dipole, this value must be scaled down by the gain of ESA dipole of 1.5. The radiating efficiency is then obtained by dividing the result by the input power (1.2 W). This gives the radiating efficiency range of $(4.4 \pm 1.7) \times 10^{-11}$. The bound (6) is $2.4 \times 10^{-11}$. Thus, the deduced radiation efficiency appears slightly higher than the CHL-derived bound. Given uncertainties in input parameters (far field onset distance, fT-level field measurements, etc.), the agreement remains reasonable.

How can we meaningfully compare the four ESEs considered here – despite vast differences in size, frequency and bandwidth? We address this by developing an ESE figure of merit (FOM) that normalizes these differences and allows for consistent, physics-based comparison across different systems.

In the deep subwavelength regime, $ka \ll 1$, and we can write $Q_{CHL} = (ka)^{-3}$. Noting that the volume of the sphere enclosing the emitter is $V = 4\pi a^3/3$, we find from (6) the limit on attainable radiation power density per unit input power:

$$p_{rad} = \frac{P_{rad}}{P_{in}} \frac{1}{V} \leq 6\pi^2 \frac{f^4}{c^3 \Delta f}, \qquad (8)$$

As seen from (8), this limit depends strongly on frequency and bandwidth. To enable ESE comparisons across frequencies and bandwidth, we introduce the ESE figure of merit:

$$FOM = \frac{P_{rad}}{P_{in}} \frac{1}{V} \frac{c^3 \Delta f}{6\pi^2 f^4}. \quad (9)$$

It quantifies how far each design lies from a theoretically attainable performance ($FOM = 1$). Table 2 lists FOM values for the four ESEs considered in this section. The FOM for LN mechanical antenna was computed using the lower estimated value for radiating efficiency, $1 \times 10^{-8}$.

Table 2. ESE FOM values

| ESE | FOM |
| --- | --- |
| ELF | 0.0064 |
| VLF | 0.048 |
| PZT mechanical | ~1 |
| LN mechanical | 0.55 |

As seen in Table 2, mechanical antennas approach the achievable limit. This conflicts with the claim of orders of magnitude improvements expected through material and design optimization made in [2]. Instead, the CHL efficiency bound establishes the fundamental constraints on achievable combinations of material properties.

In the next section, we turn to quantum emitters and examine how CHL would dictate bounds on their intrinsic properties – lifetimes of excited states and transition dipole moments. As discussed earlier, it is a hypothesis that a quantum emitter undergoing spontaneous emission is subject to CHL; the analysis below provides a direct framework for testing this hypothesis.

### IV. CHL-derived bounds on atomic transition rates

Consider a two-level atomic system with the transition frequency $f$ and linewidth $\Delta f$. From the energy-time uncertainty principle

$$\Delta E \Delta t \geq \frac{h}{4\pi}, \quad (10)$$

where $h$ is the Planck constant and $\Delta E = h \Delta f$, we find the upper bound on the bandwidth $Q$-factor:

$$Q_{bw} = \frac{f}{\Delta f} \leq 4\pi f \Delta t. \quad (11)$$

A spontaneously emitting atom is a CHL-compliant emitter, and its spectral line is Lorentzian. Therefore, its bandwidth $Q$-factor is also bounded from below by the CHL

$$4\pi f \Delta t \geq Q_{bw} \geq Q_{CHL}. \quad (12)$$

This yields a bound on the natural lifetime of the excited state

$$\Delta t \geq \frac{Q_{CHL}}{4\pi f}. \quad (13)$$

For electric dipole transitions, this can be converted to a bound on the transition dipole moment $d$ via the Einstein's $A$-coefficient [16]:

$$A = \frac{2\omega^3}{3\pi\varepsilon_0 hc^3} |d|^2. \quad (14)$$

Since $A = 1/\Delta t$, from (13) and (14), we find

$$|d| \leq \sqrt{\frac{3\varepsilon_0 h c^3}{4\pi^2 f^2} \frac{1}{Q_{CHL}}}. \quad (15)$$

A ground state of an atom and its first excited state can generally be treated as a two-level system if other energy levels lie sufficiently far so that they do not overlap the transition linewidth. Hydrogen ($1S$-$2P$) and alkali $D$-lines ($n^2S_{1/2}$-$n^2P_{1/2}$, $D_1$ line, and $n^2S_{1/2}$-$n^2P_{3/2}$, $D_2$ line) satisfy these assumptions well.

Table 3 compares the bounds (13) and (15) with the corresponding measured values for hydrogen, rubidium ($Rb$) and cesium ($Cs$) atoms. The actual lifetimes and transition dipole moments for $Cs$ and $Rb$ are taken from [17], while those for hydrogen are from [18]. The radii of the enclosing spheres for $Cs$ and $Rb$ were computed using the Alkali Rydberg Calculator (ARC) software [19] as

$$a(nP_j) = \sqrt{\langle \Psi_{nP_j} | r^2 | \Psi_{nP_j} \rangle}. \quad (16)$$

For hydrogen atom, the integral (16) is evaluated in closed form to be $\sqrt{3}$ in atomic units [18].

Table 3. Bounds on lifetime and transition dipole moments compared to their actual values. Chu radius and dipole moments are in atomic units. Lifetimes are in ns.

| Element | Transition | Chu radius a | Bound on lifetime | Actual lifetime | Bound on d | Actual d |
|---|---|---|---|---|---|---|
| H | $2P \rightarrow 1S$ | 1.73 | 0.01 | 1.6 | 9.61 | 0.745 |
| Cs | $D_1$ | 8.18 | 8.46 | 34.79 | 6.46 | 3.19 |
| Cs | $D_2$ | 8.42 | 6.38 | 30.41 | 6.92 | 4.48 |
| $^{87}Rb$ | $D_1$ | 7.72 | 6.29 | 27.68 | 6.29 | 2.99 |
| $^{87}Rb$ | $D_2$ | 7.82 | 5.61 | 26.24 | 6.47 | 4.23 |

## V. Conclusions

Despite a century of development, practical ESA sizes remain unchanged since Marconi reflecting the fundamental Chu-Harrington limit on the ESA performance. Recently proposed naturally resonant electrically small emitters, such as mechanical resonators and quantum emitters, eliminate matching circuits required by ESAs to compensate for large capacitive reactance. These matching circuits exacerbate miniaturization challenges, as the required inductor size and associated conductor losses grow rapidly with decreasing electrical size. Mechanical and quantum emitters exhibit different scaling of the resonance costs (size, loss) with the electrical size and may offer improvement over ESAs. This has been demonstrated for mechanical antennas in [1] and [2] and theorized for quantum emitters in [4] and [5].

Does this solve the antenna miniaturization problem? This paper shows mechanical antennas achieve radiating power densities approaching theoretical limit for CHL-compliant emitters, though absolute values remain low. Further gains require relaxing CHL assumptions themselves - via classical methods (non-Foster networks, nonlinear parametric antennas), or quantum effects (superradiance, quantum coherence control).

### Acknowledgement

This work was partially done under NASA contract 80NSSC24CA108. The author is grateful to Dr. D. Rowling at NASA for fruitful discussions.

### Conflict of Interest

The authors have no conflicts of interest to declare.